# Massive and massless Dirac fermions in $Pb_{1-x}Sn_xTe$ topological crystalline insulator probed by magneto-optical absorption

B.A. Assaf,[1]* T. Phuphachong,[2]* V.V. Volobuev,[3,4] A. Inhofer,[2] G. Bauer,[3] G. Springholz,[3] L.A. de Vaulchier,[2]  Y. Guldner[2]

[1]Département de Physique, Ecole Normale Supérieure, CNRS, PSL Research University, 24 rue Lhomond, 75005 Paris, France

[2]Laboratoire Pierre Aigrain, Ecole Normale Supérieure, CNRS, PSL Research University, Université Pierre et Marie Curie, Université Denis Diderot, 24 rue Lhomond, 75005 Paris, France

[3]Institut für Halbleiter und Festkörperphysik, Johannes Kepler Universität, Altenberger Strasse 69, 4040 Linz, Austria

[4]National Technical University, Kharkiv Polytechnic Institute, Frunze Ste. 21, 61002 Kharkiv, Ukraine

*These authors contributed equally to this work

Abstract: Dirac fermions in condensed matter physics hold great promise for novel fundamental physics, quantum devices and data storage applications. IV-VI semiconductors, in the inverted regime, have been recently shown to exhibit massless topological surface Dirac fermions protected by crystalline symmetry, as well as massive bulk Dirac fermions. Under a strong magnetic field (B), both surface and bulk states are quantized into Landau levels that disperse as $B^{1/2}$, and are thus difficult to distinguish. In this work, magneto-optical absorption is used to probe the Landau levels of high mobility Bi-doped $Pb_{0.54}Sn_{0.46}Te$ topological crystalline insulator (111)-oriented films. The high mobility achieved in these thin film structures allows us to probe and distinguish the Landau levels of both surface and bulk Dirac fermions and extract valuable quantitative information about their physical properties. This work paves the way for future magnetooptical and electronic transport experiments aimed at manipulating the band topology of such materials.

Topological crystalline insulators (TCI) are a novel class of materials that host an even number of Dirac surface states at points that are mirror symmetric with respect to a certain crystallographic plane.[1] The (100) and (111) surfaces of rocksalt IV-VI semiconductors $Pb_{1-x}Sn_xTe$ and $Pb_{1-x}Sn_xSe$ (Fig. 1(a))  were shown to possess such Dirac surface states.[2,3,4,5,6,7,8,9,10]  In $Pb_{1-x}Sn_xTe$ and $Pb_{1-x}Sn_xSe$, a band inversion occurs at the L-points of the Brillouin zone, as the Sn content is increased.[11,12,13] Since the L-points are mirror symmetric with respect to the (110) diagonal planes, such a band inversion results in the emergence of topological surface states (TSS) at 4 different points on the (100)[3,4,5] and the (111) surfaces[7,8,9,14,15]. IV-VI semiconductors are thus 4-fold degenerate TCI where topology is governed by the symmetry of the crystal. As such, TCI hold a great potential for tunable Dirac electronics,[16,17,18,19] stemming from the inherent sensitivity of band topology on the crystal structure,[2,20] as well as from the highly mobile character of Dirac Fermions.[21,22] A complete understanding of the behavior of surface Dirac fermions in such materials and the ability to reliably distinguish them from the bulk carriers is, however, a necessary prerequisite to their development and implementation in potential devices. [16,23,24]

As in most other topological materials, the bulk states may also contribute in magnetotransport experiments as well as in optics. As shown in Fig. 1(b), the bulk carriers in (111) oriented $Pb_{1-x}Sn_xTe$ populate 4 ellipsoidal carrier pockets, one of which is a longitudinal pocket whose major axis is parallel to the [111] direction, while the other three have their major axes tilted by ±70.5⁰ with respect to the

[111] direction. In total, one has to deal with a complex Fermi surface that comprises a longitudinal bulk ellipsoidal valley, three tilted oblique bulk valleys, a $\bar{\Gamma}$-point surface Dirac valley, and three $\bar{M}$-point surface Dirac valleys. The bulk states are also expected to be Dirac-like,[25] and may result in carrier mobilities that could be as high as what is expected for the topological surface states. Hence, in order to reliably identify the topological surface states and study them with conventional transport and optical probes, one needs to be able to understand and rule out contributions from bulk states that may contribute similar, if not identical signatures. Additionally, previous magnetooptical[26,27] and transport studies[28,25,29] have proven difficult the task of probing, identifying and assigning the Landau levels of topological materials. This is mainly due to the fact that the Fermi level in such systems is pinned to the bulk states, and the mobility is limited, hence requiring fields exceeding 15T to achieve clear Landau quantization.[30]

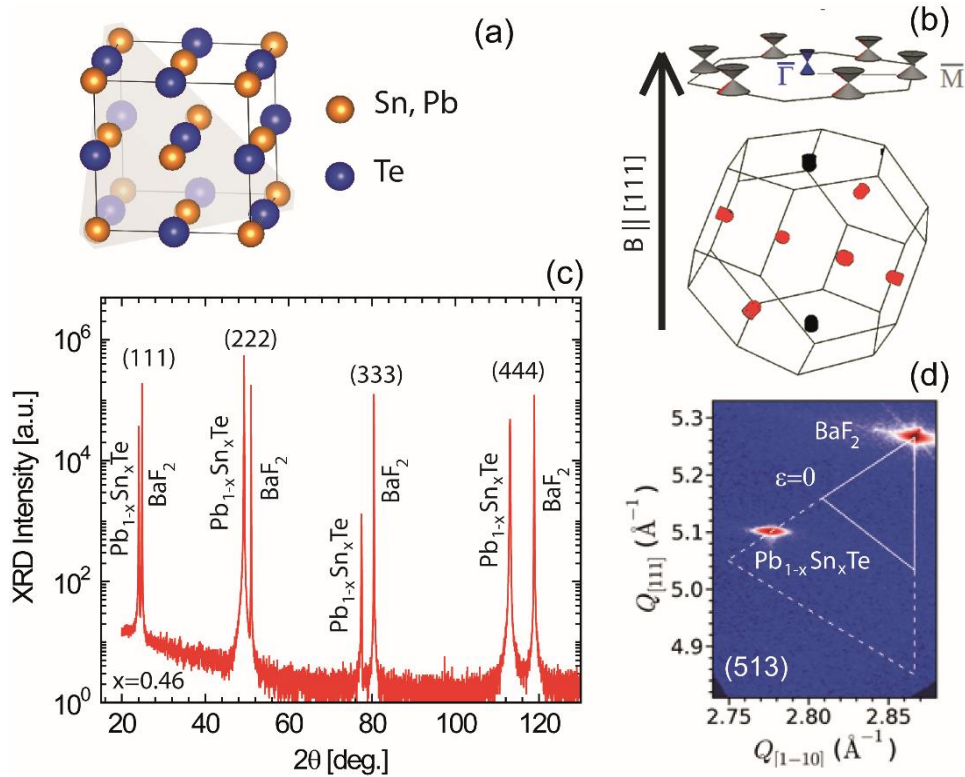

Figure 1. The structural properties of $Pb_{1-x}Sn_xTe$ (111) films and sketch of their Brillouin zone. (a) Rocksalt (Fm-3m) crystal structure of $Pb_{1-x}Sn_xTe$ highlighting a (111) Bragg plane in grey. (b) Sketch of the first Brillouin zone of $Pb_{1-x}Sn_xTe$ rotated so that a (111) plane points towards the top. The longitudinal bulk valleys are shown in black, and the oblique bulk valleys are shown in red. A sketch of the (111) topological surface Dirac cones is shown on top. The $\bar{\Gamma}$ Dirac cone shown in blue has a larger Fermi velocity than the three $\bar{M}$ Dirac cones. (c) Symmetric x-ray diffraction scan along [111] (using Cu-K$\alpha_1$ radiation) of the $Pb_{1-x}Sn_xTe$ sample S1 with x = 0.46 showing the epitaxial structure of the layer with respect to the substrate. (d) Reciprocal space map taken around the asymmetric (513) Bragg reflection, evidencing that the epilayer is strain free, i.e. fully relaxed.

In this work, we performed detailed magneto-optical absorption measurements to map out the Landau level (LL) spectrum of the bulk and surface bands of high mobility (111) epitaxial $Pb_{1-x}Sn_xTe$ (x=0.45-0.47) films grown by molecular beam epitaxy (MBE). By lightly doping $Pb_{1-x}Sn_xTe$ with Bi (about $10^{19}cm^{-3}$),[31] we are able to compensate residual defect doping that results from (Pb,Sn) vacancies, and achieve a low carrier density without compromising the mobility. We can, hence, obtain bulk carrier densities that are close to 1x10$^{18}$ cm$^{-3}$ and mobilities of 10000 cm$^2$/Vs. This results in a Landau quantization at relatively low magnetic fields (1.5T) and allows us to reliably map the LL spectrum of both bulk and surface states.

Magneto-optical measurements reveal a number of strong interband Landau level transitions that are well described by a massive Dirac dispersion model.[25] These transitions are associated with two types of bulk ellipsoidal Fermi surfaces – a longitudinal carrier valley and three-fold degenerate oblique valleys shown in Fig. 1(b). After reliably mapping out all bulk contributions, we are able to identify a cyclotron resonance feature pertaining to a massless Dirac state having a Fermi velocity v$_f$=7.3x10$^5$m/s, attributed to the $\bar{\Gamma}$-point Dirac cone. This is reproduced in two samples having slightly different carrier densities. Our results are in agreement with previous studies on the bulk bands in SnTe[32], Pb$_{1-x}$Sn$_x$Te[33], and PbTe[34,35,36] and recent calculations of the band structure of the (111) surface states in TCI systems.[9,14]

# Results

## Growth and characterization

High mobility Pb$_{1-x}$Sn$_x$Te films are grown by molecular beam epitaxy (MBE) on cleaved (111) BaF$_2$ substrates, in a Varian Gen II system with a base pressure of 10$^{-10}$ mbar. Compound PbTe and SnTe effusion cells are employed to control the composition of the layer, through the PbTe-to-SnTe flux ratio that is determined using a quartz microbalance moved into the substrate position. For the present investigations, the Sn composition of the layers was fixed to x$_{Sn}$ = 0.46±0.01, which is well in the non-trivial TCI regime. The thickness was fixed at 2µm. Bismuth n-type doping was employed to compensate the intrinsic p-type carrier concentration (typically p>10$^{19}$cm$^{-3}$) arising in Pb$_{1-x}$Sn$_x$Te from native Sn and Pb vacancies. Bi was supplied from a compound Bi$_2$Te$_3$ effusion cell. When substitutionally incorporated on group IV lattice sites, Bi acts as singly charged donor and thus, can compensate the p-type background concentration. From a sample series with varying Bi doping levels, selected samples with carrier concentration around 1 x 10$^{18}$ cm$^{-3}$ and carrier mobilities of 10000 cm$^2$/Vs are chosen for further investigation. Apart from the electrical measurements, all samples are characterized by high resolution X-ray diffraction (XRD) shown in Fig. 1(c ,d) and atomic force microscopy (AFM, see supplement). From XRD (Fig. 1(c)), a clear Pb$_{1-x}$Sn$_x$Te {111} Bragg series can be observed, evidencing a perfect epitaxial relationship between layer and substrate. The layers are fully relaxed with practically zero residual strain as shown by the reciprocal space map around the asymmetric (513) Bragg reflection (Fig. 1(d)), where the Pb$_{1-x}$Sn$_x$Te peak is found to be located exactly on the zero strain line (ε=0). The composition of the layers derived from XRD using the Vegard's law agrees within ±1% with the nominal value.

## Magnetooptics-Bulk States

Infrared magneto-optical spectroscopy experiments are performed on large pieces of two samples (S1 and S2), at 4.5K and up to B=15T. The samples are exposed to radiation from a mercury lamp in the Faraday geometry, with the magnetic field parallel to the [111] direction. The transmitted signal is then collected using a Si composite bolometer and analyzed by a Fourier transform spectrometer. Figures 2(a, b) show typical infrared transmission spectra taken at different magnetic fields between 55 and 400 meV in both samples. A number of strong absorption minima that disperse with increasing magnetic field can be clearly seen. The strongest series marked by black dots is associated with the bulk longitudinal valley. Several other transitions marked by the red circles, can also be resolved and are attributed to bulk oblique valleys. Note that the mere fact that a strong and clear modulation is seen versus energy at fields as low as 1.5T is unambiguous evidence of the high mobility and low carrier density of the films.

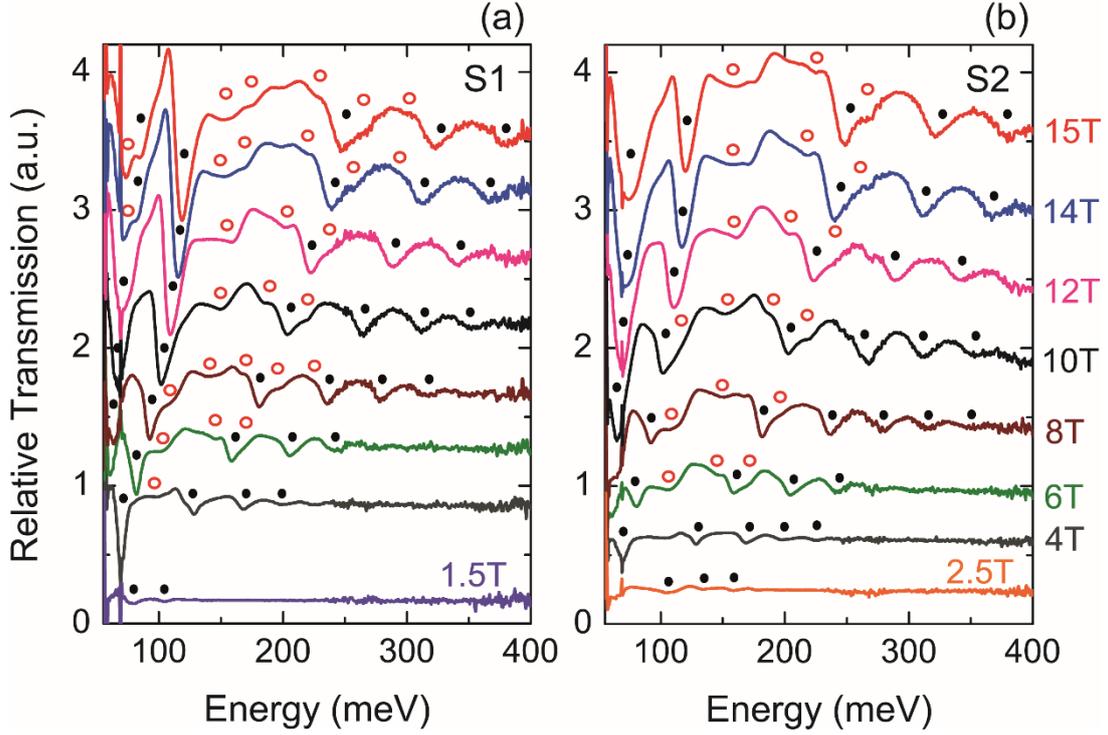

**Figure 2. Magnetooptical spectra of Pb$_{0.54}$Sn$_{0.46}$Te (111) films.** (a) and (b) Magneto-optical transmission spectra (calculated by dividing the raw spectra by the spectrum at B=0T) as a function of energy, plotted for magnetic fields between 1.5T and 15T in sample S1 and 2.5T and 15T in S2. The full black circles and empty red circles respectively denote absorption minima corresponding to the longitudinal and oblique valleys. The curves are shifted vertically for clarity purposes. The slope in the baseline of the spectra originates from the response of the detector to the applied magnetic field. Its impact on the position of the minima is negligible, since it is smaller than our reported error bars.

Figures 3(a,b) respectively show the detailed analysis performed for S1 and S2, whereby absorption energy minima are identified (Fig. 2(a, b)) and then plotted as a function of magnetic field in order to construct a Landau fan diagram. A massive Dirac model is then used to fit the data for both types of valleys. [25,37,38] The LL energies at $k_z = 0$ (B and z||[111]) are sufficient to describe the magnetooptical absorption spectra since the joint density of states is optimal for $k_z = 0$. Taking the zero energy at the midgap, the LL energies given by the massive Dirac model are:

$$E_N^{c,v} = \pm\sqrt{\Delta^2 + 2e\hbar v_f^2 NB} \qquad (1)$$

The ± sign refers to the conduction ($E_N^c$) and valence ($E_N^v$) band levels respectively and Δ is equal to half of the band gap E$_g$, $\Delta = \frac{E_g}{2}$. The mass of the Dirac Fermions is then given by $m_D = \frac{\Delta}{v_f^2}$, where v$_f$ is the Fermi velocity. N denotes the Landau index, e is the fundamental electronic charge, and ℏ is the reduced Planck constant. The absorption energies for interband transitions are then given by:

$$E_{N-1}^c - E_N^v = \sqrt{\Delta^2 + 2e\hbar v_f^2 (N-1)B} + \sqrt{\Delta^2 + 2e\hbar v_f^2 NB} \qquad (2)$$

The cyclotron resonance (CR) as a function of field satisfies the intraband equivalent:

$$E_0^v - E_1^v = -\Delta + \sqrt{\Delta^2 + 2e\hbar v_f^2 B} \qquad (3)$$

Note that a similar expression can also be derived from a two-band **k.p** model that includes spin degeneracy.[36,37,39,40,41] For the Sn content that we considered in this work (x≈0.46),[42] a two-band approach is justified, since the conduction ($L^{6+}$) and valence ($L^{6-}$) bands are mirror images of each other and the far-band contributions[13,38] can be neglected for an energy gap of 30meV. In this case, the LL spin splitting is known to be almost equal to the cyclotron energy, so that the $N^{th}$ LL of one spin component coincides with the $(N+1)^{th}$ LL of the other.[36,43,13] This also follows directly from the Dirac Hamiltonian[37] and is even well known to be valid in general for Dirac electron propagating in vacuum. The spin degeneracy will therefore not give additional transmission minima as discussed in more detail in the supplement.

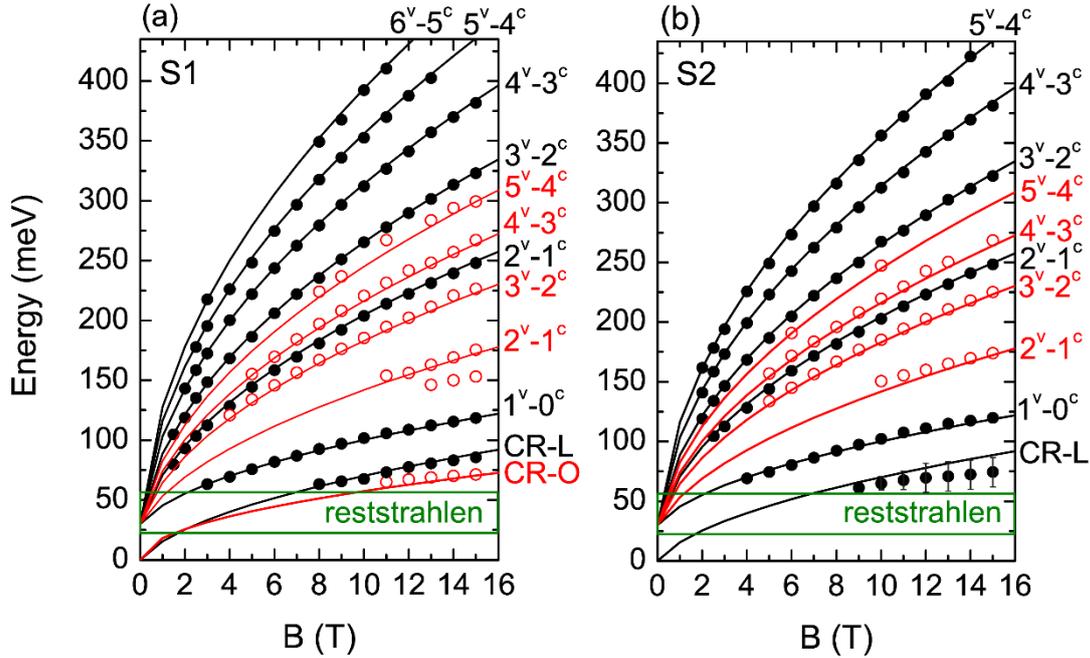

Figure 3. Experimental and theoretical Landau fan diagrams of $Pb_{0.54}Sn_{0.46}Te$ (111) films. Experimental Landau fan diagram for the longitudinal (full black circles) and oblique (empty red circles) bulk valleys for S1 (a) and S2 (b). Solid lines are theoretical curves obtained from a massive Dirac model (Eq. (2,3)) with $v_f = 7.3 \times 10^5$ m/s for the longitudinal valley (black lines) and $v_f = 5 \times 10^5$ m/s for the oblique valleys (red lines) and Δ≈15meV. The $BaF_2$ substrate reststrahlen band (22 to 55meV) is shown in green. CR denotes the cyclotron resonance absorption. Interband absorptions $N^v – (N-1)^c$ are denoted by their respective Landau index N and their band index (c for the conduction band and v for valence band). The CR-data points in S1 are extracted using a multi-peak fitting method described later in the text. CR-L and CR-O respectively denote the CR originating from the longitudinal and the oblique valleys.

The black lines in Fig. 3 are the calculated magneto-optical transition energies for the longitudinal valley using the massive Dirac model described earlier. A very good agreement between the theory and the experimental interband data is obtained for $v_f=(7.3\pm0.1)\times10^5$ m/s and Δ=15±3meV for both samples. The energy gap $E_g=2\Delta\approx30\pm6$meV agrees with the value expected for $Pb_{0.54}Sn_{0.46}Te$. The Fermi velocity is also, in excellent agreement with the **k.p** matrix element determined in ref. 13 (see supplement). The band edge mass is found to be equal to $0.005 m_0$. The longitudinal inter-LL transitions $1^v – 0^c$ and $2^v – 1^c$, where v and c respectively denote the valence and conduction band level, are measured down to B≈4T and 2T respectively, indicating a Fermi energy of about $E_F \approx 40\pm5$meV below the valence band edge of the longitudinal valley.

The additional minima shown as open red circles in Figures 2(a, b) are associated with carriers in the oblique valleys. As mentioned before, the Fermi surface of (111) oriented $Pb_{1-x}Sn_xTe$ is highly anisotropic.[35,36,34] The oblique valleys result in an anisotropy factor K=10, defined as the square of the ratio of the maximum and minimum cross-sectional areas of the 3D Fermi surface (Fig. 1(b)). This agrees with previous studies on the (Pb,Sn) chalcogenides.[32,35,34,36]

The red lines in Fig. 3 are the calculated magneto-optical absorption energies for the oblique valleys using the massive Dirac model. A good agreement is found between the data and the model, for a Fermi velocity v'$_f$=5x10$^5$ m/s for both samples. This corresponds to a ratio of 1.46 between the Fermi velocity of the longitudinal and the oblique valleys, in agreement with the expected Fermi surface anisotropy (see supplement). For Δ=15meV this yields a mass equal to 0.011m$_0$. The transition $3^v - 2^c$ is measured down to B=4T, indicating a Fermi energy E$_F$≈50±5meV below the valence band edge of the oblique valley for both S1 and S2. Note that it is not surprising that the Fermi level be slightly different in different valleys, as the $Pb_{1-x}Sn_xTe/BaF_2$(111) system is known to exhibit a strong thermal expansion mismatch at low temperature, that may shift the oblique bands up in energy with respect to the longitudinal bands. This has been observed in IV-VI epilayers and quantum wells grown on $BaF_2$(111).[44,45,46]

Note that the CR splitting observed in S1 above 11T in Fig. 2(a), is due to the simultaneous presence of both the longitudinal and the oblique CR lines. The critical magnetic field, above which the ($1^v$-$0^v$) oblique CR line is observed, is in agreement with the energy at which the N=1 LL crosses the Fermi level. We are, however, unable to resolve both CR lines in S2. The average peak position is thus reported with a large error bar that takes into account the broadening of the peak as a result of the presence of two transitions at those energies.

Magnetooptics – Surface states

We now turn our attention to a CR feature that seems to evade the expected physics of the bulk longitudinal and oblique valleys. Figures 4(a, b) present a zoomed in view of the spectra at high magnetic fields (11-15T) in the range 55-150 meV. Besides the main absorptions associated with the CR and the first interband transition of the longitudinal bulk valley, an absorption indicated by the blue arrows is measured.

This absorption (blue points in Fig. 4(c,d)) is interpreted as the CR of the topological surface states (CR-TSS) at the $\bar{\Gamma}$ point. The blue lines are the calculated transition energies for a massless Dirac model using the same Fermi velocity as the bulk longitudinal valley, v$_f$=7.3x10$^5$ m/s, $E_N^{c,v} = \pm\sqrt{2e\hbar v_f^2 NB}$. This gives:

$$E_{CR-TSS} = \sqrt{2e\hbar v_f^2 B} \qquad (4)$$

Good agreement is again found between the model and the experimental data for the CR-TSS feature as shown by the blue line in Figure 4(c, d). Moreover, the fact that we do not observe any CR-TSS transitions below 10T in S1 (12T in S2) agrees with the estimated Fermi levels for bulk bands (40±5meV below the band edge), as accordingly the TSS Fermi level would be close to 60meV below the Dirac point. This fact actually reinforces our interpretation. Note, however, that the TSS interband transitions cannot be resolved in our experiments because they are located at the same energy as the intense interband transitions of the longitudinal bulk valley. Indeed, the LL energies of the TSS $\sqrt{2e\hbar v_f^2 NB}$ coincide with those of the massive bulk fermions $\sqrt{\Delta^2 + 2e\hbar v_f^2 NB}$ when $\sqrt{2e\hbar v_f^2 NB} \gg \Delta = 15 meV$. This is the case in our measurements for N≥1.

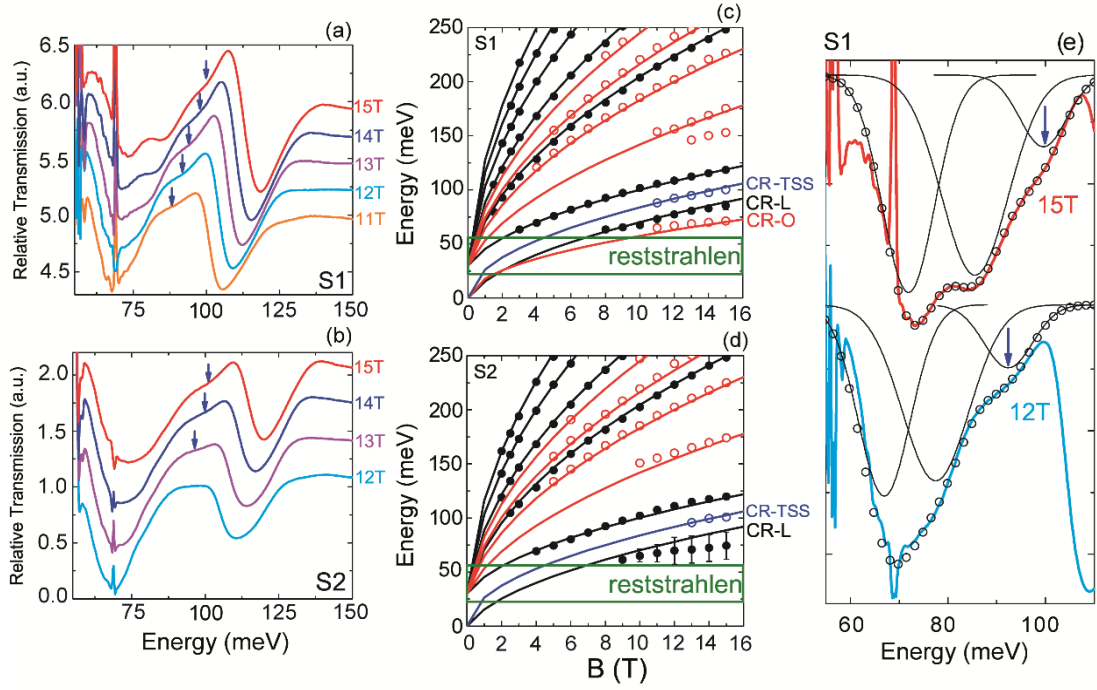

**Figure 4. The cyclotron resonance of the $\bar{\Gamma}$ point topological surface state.** Absorption minima at low energy for both S1 (a) and S2 (b) showing the cyclotron resonance feature attributed to the topological surface states (CR-TSS) between the stronger bulk absorption minima. Landau fan diagrams for S1 (c) and S2 (d) including the feature interpreted as the CR-TSS in blue. Solid blue lines are obtained by theoretically calculating the CR transition of massless Dirac Fermions having $v_f$ = 7.3x10$^5$m/s. The error bars given in (d) take into account the broadening of the bulk CR minimum seen in (b). (e) Multi-peak fitting in S1 for spectra taken at 15T (red data) and 12T (blue data). Three Gaussian curves resulting from the fitting procedure are shown. The resultant sum of the three peaks (empty black circles) gives an excellent fit to the data at both fields. The blue arrow refers to the peak attributed to the CR-TSS transition. The bulk-CR data points in S1 are also extracted using this method.

Due to the apparent low intensity of the CR-TSS in the raw signal, we have used a multi-peak fitting scheme to identify the position of the transition minima in each case. This procedure is shown in Fig. 4(e). We can clearly identify three peak features; two strong features resulting from the CR of the bulk states as well as a smaller one indicated by the blue arrow that we attribute to the CR-TSS. The feature marked by the blue arrow is essential to explain the dip observed in the raw signal, and is sufficiently large in intensity to be reliably accounted for in the analysis. At 15T we find the CR-TSS at 100±2meV, whereas at 12T it is located at 92±2meV. The blue arrows and the CR data points in Fig. 3 and 4 are identified and placed using this method.

## Discussion

In sum, we are able to map out the band structure of Bi-doped Pb$_{0.54}$Sn$_{0.46}$Te in the vicinity of the band gap. Our interpretation is summarized in Fig. 5. The longitudinal bulk valley is found to satisfy a massive Dirac Hamiltonian with a Fermi velocity $v_f$=(7.3±0.1)x10$^5$m/s and an effective mass equal to m*=0.005m$_0$. This Fermi velocity is in excellent agreement with the **k.p** matrix element determined in ref. 13 (see supplement). The bulk oblique valleys satisfy that same massive Dirac Hamiltonian with

v′$_f$=(5.0±0.1)x10$^5$m/s and an effective mass m'*=0.011m$_0$. Finally, a CR that satisfies a massless Dirac model with v$_f$=(7.3±0.1)x10$^5$m/s is observed and attributed to the TSS $\overline{\Gamma}$-point Dirac cone. We did not see any signal from LL pertaining to the $\overline{M}$-point Dirac cones. The reason behind this might be the fact that the Fermi velocity for $\overline{M}$-point Dirac cones coincides with that of the oblique bulk valleys. In that case, the interband transitions of the $\overline{M}$-point Dirac cone will overlap with those of the oblique bulk valleys. The CR signal for $\overline{M}$-point Dirac cones is likely to occur at energies similar to where the bulk-CR lines were observed. It is also expected to be weaker in intensity; we are thus not able to resolve it.

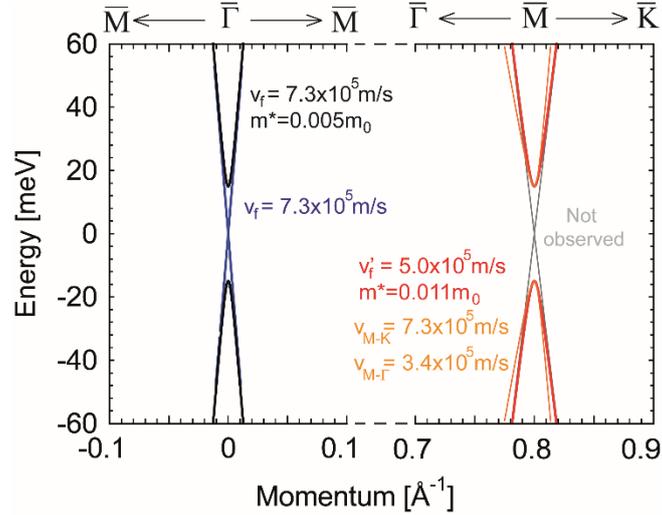

**Figure 5. Summary of band parameters extracted in this study.** Band structure plotted using the parameters extracted for the bulk massive Dirac fermions (black, red and orange) and the topological massless Dirac fermions (blue).

Finally, one has to keep in mind that the Fermi velocity of the oblique valleys is expected to be anisotropic.[9] This is due to the fact that the oblique valleys are tilted by 70.5° with respect to the [111] direction and the applied magnetic field. Our experimental value of v′$_f$ = (5.0±0.1)x10$^5$m/s is thus an effective result given by $v'_f = \sqrt{v_{\overline{M}-\overline{K}} v_{\overline{M}-\overline{\Gamma}}}$. Here $v_{\overline{M}-\overline{K}}$ denotes the Fermi velocity along the $\overline{M}-\overline{K}$ direction, expected to be almost equal to that measured at the $\overline{\Gamma}$-point – (7.3±0.1)x10$^5$m/s.[9] One can thus extract $v_{\overline{M}-\overline{\Gamma}}$, the Fermi velocity along the $\overline{M}-\overline{\Gamma}$ direction. We get $v_{\overline{M}-\overline{\Gamma}} \approx (3.4\pm 0.2)\times 10^5 m/s$. The details of this geometric argument are discussed in the supplement.

In conclusion, we have mapped out the Landau level spectrum of high mobility Bi-doped Pb$_{0.54}$Sn$_{0.46}$Te (111) epilayers grown on BaF$_2$. The high mobility and low carrier density achieved in these films lead to a Landau quantization at about 1.5T. This allows us to reliably map out the LL at low energies. The bulk longitudinal and oblique valleys of Pb$_{0.54}$Sn$_{0.46}$Te can be reliably interpreted by a massive Dirac fermion model. The cyclotron resonance of the topological surface $\overline{\Gamma}$-point Dirac cone is also revealed above 10T. Its dispersion is characteristic of massless Dirac fermions having a Fermi velocity v$_f$ = 7.3x10$^5$m/s. The interband transitions corresponding to the TSS cannot be resolved as they overlap those of the bulk longitudinal valley. Our results provide vital information about the effective masses and Fermi velocities of different bands in Pb$_{0.54}$Sn$_{0.46}$Te, most notably the $\overline{\Gamma}$-Dirac point, and will be of great use to future transport experiments studying quantum oscillations and coherent transport.

## Methods

Bi-doped $Pb_{1-x}Sn_xTe$ films were grown by molecular beam epitaxy in a Varian Gen II MBE system on cleaved $BaF_2$(111) substrates. High purity PbTe, SnTe and $Bi_2Te_3$ source materials were used. High resolution X-ray diffraction measurements were then performed using Cu-K$\alpha_1$ radiation in a Seifert XRD3003 diffractometer, equipped with a parabolic mirror, a Ge(220) primary beam Bartels monochromator and a Meteor 1D linear pixel detector. Preliminary transport measurements were performed at 77K in order to determine the carrier density and mobility. Further transport characterization was performed at 2K and up to 8T in an Oxford Instruments 1.5K/9T cryostat. Magneto-optical absorption experiments were performed in an Oxford Instruments 1.5K/15T cryostat at 4.2K. Spectra were acquired using a Bruker Fourier transform spectrometer.

## Additional Information

### Acknowledgements

B.A.A. acknowledges support from the LabEx ENS-ICFP: ANR-10-LABX-0010/ANR-10-IDEX-0001-02 PSL. V.V.V., G.B., and G.S. acknowledge support from Austrian Science Fund IRON SFB F2504-N17.

### Author contributions

B.A.A. and Y.G. conceived the experiment. T.P., L.A.V. and Y.G. performed the spectroscopy experiments. B.A.A., T.P., L.A.V. and Y.G. analyzed the resulting spectra with the assistance of G.B. B.A.A. and A.I. performed high-field transport characterization. V.V.V. and G.S. grew and characterized the thin film samples. B.A.A. wrote the manuscript with the assistance of G.B., G.S., L.A.V. and Y.G. All authors reviewed and discussed the final results.

### Competing financial interest

The authors declare no competing financial interest.

# Supplementary material for Massive and Massless Dirac fermions in Pb$_{1-x}$Sn$_x$Te topological crystalline insulator probed by magneto-optical absorption

B.A. Assaf, T. Phuphachong, V.V. Volobuev, A. Inhofer, G. Bauer, G. Springholz, L.A. de Vaulchier, Y. Guldner

## 1. Transport parameters and structural characterization

The Hall carrier density and mobility were measured in both samples at 2K. Hole-type carriers were found to contribute for both. The carrier densities deduced from a high-field Hall measurement (up to 8T) were measured to be slightly different in the two samples: p=1.0×10$^{18}$cm$^{-3}$ for S1 and p=1.2×10$^{18}$cm$^{-3}$ for S2. The Hall mobility was found to be equal to 4800 cm$^2$/Vs in S1 and 10500 cm$^2$/Vs in S2. Let us keep in mind, however, that this is an effective mobility that is averaged over all band and valley contributions. The carrier mobility in the longitudinal bulk valley which depends only on the transverse effective mass and in the Dirac cones may thus be larger than these reported values.

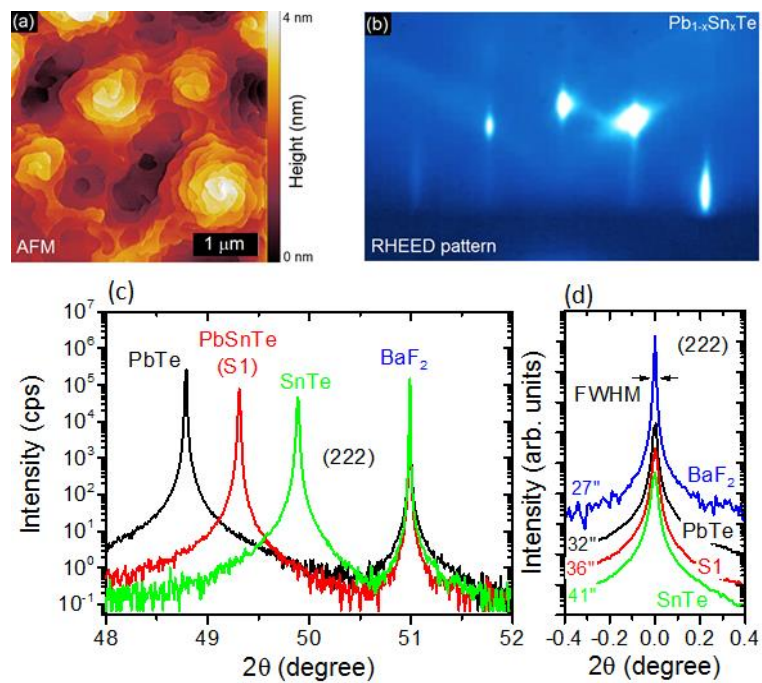

Supplementary Figure 1. (a) Atomic force microscopy (AFM) image of a 2µm Pb$_{1-x}$Sn$_x$Te epilayer on BaF$_2$(111) substrate. (b) Reflection high energy electron diffraction pattern obtained in-situ during the growth of a Pb$_{1-x}$Sn$_x$Te layer on a BaF$_2$ (111) substrate, showing a series of equally spaced streaks, evidence of a high quality (111) oriented epitaxial layer. (c) High resolution x-ray diffraction pattern recorded in 4-crystal geometry on the Pb$_{0.54}$Sn$_{0.46}$Te epilayer S1 and pure PbTe and SnTe epilayers grown under the same conditions on BaF$_2$ (111) substrates. From the peak position, the lattice constants and thus, the chemical composition is derived using Vegard's law. (d) Comparison of the FWHM of the (222) diffraction peak in all three samples. The practically identical FWHM (shown on the left side of the curves) found in the three samples proves the absence of any compositional alloy fluctuations in the ternary Pb$_{0.54}$Sn$_{0.46}$Te layer.

The samples are additionally characterized by AFM as shown in Suppl. Fig. 1(a). A number of screw-type dislocations can be seen in the AFM image. The surfaces, however, exhibit only single monolayer steps around growth spirals induced by the dislocations, originating from the lattice mismatch with the BaF$_2$ substrates (Δa/a = 3.07 %). Locally, terraces that are several hundreds of nanometers in length and width are seen. Finally, a RHEED pattern of the Pb$_{1-x}$Sn$_x$Te layer is shown in Suppl. Fig. 1 (b). A series of streaks is seen in the figure evidencing the high crystalline quality of the film. Additionally, a through XRD comparative study is performed on three samples, S1, a PbTe epilayer and a SnTe epilayer, all grown under identical conditions on BaF$_2$ (111). The (222) Bragg peak for the three samples is shown in Suppl. Fig. 1(c). The peak position changes upon Sn doping of PbTe due to the changing lattice constant, as dictated by Vegard's law. We then compared the FWHM of the (222) Bragg peaks in Suppl. Fig 1(d). It is evidently clear that the FWHM (36'') of the (222) peak for sample S1 is identical to that of SnTe (41'') and PbTe (32''), even in the presence of Bi-doping. We can thus rule out the existence of compositional inhomogeneities and concentration gradients in the sample, therefore confirming that a single phase is successfully formed in S1.

## 2. Equivalence between the two-band k.p model and the massive Dirac model in Pb$_{1-x}$Sn$_x$Te

### The k.p and Dirac Hamiltonians in IV-VI semiconductors

According to previous studies [1,2,3,4,5,6,7] a six-band k.p Hamiltonian provides a good description of the electronic band structure of PbTe, and consequently other IV-VI semiconductors that crystallize in the rocksalt structure with direct energy gaps at the L-points of the Brillouin zone. For narrow (nearly zero) gap alloys of SnTe and PbTe such as Pb$_{0.54}$Sn$_{0.46}$Te, it can, however, be argued that a two-band description is sufficiently accurate. A two-band k.p Hamiltonian matrix is given by:

$$\begin{pmatrix} \frac{E_g}{2}-E & 0 & \frac{\hbar}{m_0}P_\parallel k_z & \frac{\hbar}{m_0}P_\perp k_- \\ 0 & \frac{E_g}{2}-E & \frac{\hbar}{m_0}P_\perp k_+ & -\frac{\hbar}{m_0}P_\parallel k_z \\ \frac{\hbar}{m_0}P_\parallel k_z & \frac{\hbar}{m_0}P_\perp k_- & -\frac{E_g}{2}-E & 0 \\ \frac{\hbar}{m_0}P_\perp k_+ & -\frac{\hbar}{m_0}P_\parallel k_z & 0 & -\frac{E_g}{2}-E \end{pmatrix} \begin{pmatrix} f_1 \\ f_2 \\ f_3 \\ f_4 \end{pmatrix} = 0 \quad (1)$$

Here E$_g$ is the band gap, $P_\parallel$ and $P_\perp$ are respectively the longitudinal and transverse momentum matrix elements. (k$_x$, k$_y$, k$_z$) are the wavevector coordinates and k$_\pm$=k$_x$±ik$_y$. m$_0$ is the electron rest mass. We define the z-axis to be oriented along the [111] direction. The magnetic field is applied in the same direction and the magneto-optical transitions occur at k$_z$=0. We thus get:

$$\begin{pmatrix} \frac{E_g}{2}-E & 0 & 0 & \frac{\hbar}{m_0}P_\perp k_- \\ 0 & \frac{E_g}{2}-E & \frac{\hbar}{m_0}P_\perp k_+ & 0 \\ 0 & \frac{\hbar}{m_0}P_\perp k_- & -\frac{E_g}{2}-E & 0 \\ \frac{\hbar}{m_0}P_\perp k_+ & 0 & 0 & -\frac{E_g}{2}-E \end{pmatrix} \begin{pmatrix} f_1 \\ f_2 \\ f_3 \\ f_4 \end{pmatrix} = 0 \quad (2)$$

This is equivalent to solving the two following decoupled matrices:

$$\begin{pmatrix} E_g/2-E & \dfrac{\hbar}{m_0}P_\perp k_- \\ \dfrac{\hbar}{m_0}P_\perp k_+ & -E_g/2-E \end{pmatrix}\begin{pmatrix} f_1 \\ f_4 \end{pmatrix}=0 \quad \text{and} \quad \begin{pmatrix} E_g/2-E & \dfrac{\hbar}{m_0}P_\perp k_+ \\ \dfrac{\hbar}{m_0}P_\perp k_- & -E_g/2-E \end{pmatrix}\begin{pmatrix} f_2 \\ f_3 \end{pmatrix}=0 \quad (3)$$

Recalling that $k_\pm = k_x \pm i k_y$, the two matrices can be rewritten in the form of a massive Dirac Hamiltonian matrix where the Fermi velocity is defined as $v_f = \dfrac{P_\perp}{m_0}$ :

$$\begin{pmatrix} E_g/2-E & \hbar v_f(k_x - i k_y) \\ \hbar v_f(k_x + i k_y) & -E_g/2-E \end{pmatrix}\begin{pmatrix} f_1 \\ f_4 \end{pmatrix}=0 \quad (4)$$

We can thus calculate $v_f$ from the value of $\dfrac{P_\perp^2}{m_0} = 2.975\,eV$ given by Bauer in ref. 5:

$$v_f = \sqrt{\dfrac{2.975 \times 1.6 \times 10^{-19}}{9.1 x 10^{-31}}} = 7.23 \times 10^5\, m/s \quad (5)$$

We measured $v_f$ = (7.3±0.1)x10$^5$m/s for the longitudinal valley, in excellent agreement with the calculated value.

Landau levels in the two-band k.p model and the massive Dirac model in Pb$_{1-x}$Sn$_x$Te

When a magnetic field is applied, $\vec{k}$ is replaced by $\vec{k} - e\vec{A}/\hbar$. According to ref. 5 Landau quantization in the two-band k.p model results in the following:

$$\begin{pmatrix} E_g/2-E & \sqrt{E_g\hbar\omega\left(n+\dfrac{1}{2}\right)+E_g\dfrac{g\mu_B B}{2}} \\ \sqrt{E_g\hbar\omega\left(n+\dfrac{1}{2}\right)+E_g\dfrac{g\mu_B B}{2}} & -E_g/2-E \end{pmatrix}\begin{pmatrix} \varphi_1 \\ \varphi_2 \end{pmatrix}=0 \quad (6)$$

Here ω is the cyclotron frequency, $g$ is the g-factor and $\mu_B$ the Bohr magneton. $\varphi_1$ and $\varphi_2$ are the harmonic oscillator wavefunctions ($\varphi_n$ and $\varphi_{n+1}$). It can also be shown that for Dirac electrons as well as in the two-band k.p approximation the spin splitting is equal to the cyclotron energy,[4,5,7] which gives:

$$\begin{pmatrix} E_g/2-E & \sqrt{E_g\hbar\omega(n+1)} \\ \sqrt{E_g\hbar\omega(n+1)} & -E_g/2-E \end{pmatrix}\begin{pmatrix} \varphi_1 \\ \varphi_2 \end{pmatrix}=0 \quad (7)$$

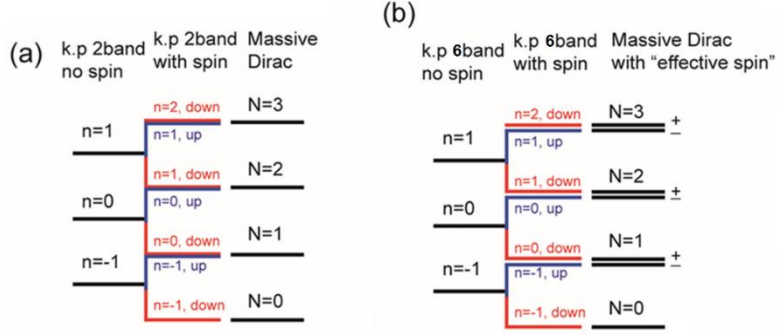

Supplementary Figure 2. Illustration of the Landau level spectrum obtained from (a) a two-band **k.p** model (and its equivalent, the Massive Dirac model) and (b) a six-band **k.p** model (and its equivalent, the Massive Dirac model that includes an "effective spin").

If we define $\Delta \equiv E_g/2$, and $v_f^2 = \Delta/m^*$, where $m^*$ is the effective mass in $\omega=eB/m^*$, and redefine our Landau index $n+1$ to be equal to N, we further simplify the notation to yield:

$$\begin{pmatrix} \Delta - E & \sqrt{2\hbar v_f^2 eBN} \\ \sqrt{2\hbar v_f^2 eBN} & -\Delta - E \end{pmatrix} \begin{pmatrix} \varphi_1 \\ \varphi_2 \end{pmatrix} = 0 \quad (8)$$

This is the Landau level matrix of a massive Dirac fermion band dispersion of gap equal to $2\Delta$. We recover the massless equivalent by setting $\Delta=0$.

The LL energies are then given by:

$$E_N = \pm\sqrt{\Delta^2 + (2\hbar v_f^2 eBN)} \quad (9)$$

Note that in both the two-band **k.p** model and the massive Dirac model, all LL are spin degenerate except the N=0 level which is non-dispersive and is comprised of a single spin component as illustrated in Suppl. Fig. 2(a).[5,8]

Far band contributions are negligible in small gap compounds

In this entire study, we have neglected the contribution of the far bands. This is justified by the fact that the band gap in $Pb_{0.54}Sn_{0.46}Te$ is about 30meV and thus favors interactions between the highest valence band and lowest conduction band. The far-band correction to the two-band **k.p** model is discussed in detail in ref. 4 and ref. 5 for trivial IV-VI semiconductors. Small corrections to the cyclotron mass and g-factor result from far-band terms in the Hamiltonian. Such far band contributions have not, however, been determined for $Pb_{0.54}Sn_{0.46}Te$. We are thus not capable of providing a detailed quantitative analysis of their impact in our case.

If we use the parameters provided for PbTe in ref. 5 for the 6-band terms, the far-band corrections can be approximated. Corrections on the order of 5% are estimated for the effective masses. The g-factor corrections are even smaller and do not exceed 2%. They result in corrections in energy that are two orders of magnitude smaller than the energies measured in the case of the interband and intraband transitions at high magnetic field. The 6-band treatment will also lift the degeneracy of the N≥1 levels, and yield an "effective spin splitting" as shown in Suppl. Fig. 2(b). Both impacts of the 6-band model are negligible within our experimental resolution.

On a separate note, far-band contributions can yield a crossing of the conduction and valence N=0 levels at very high magnetic fields, for negative band gap (topologically non-trivial) alloys such as $Pb_{0.54}Sn_{0.46}Te$.

This has been measured in $Pb_{1-x}Sn_xSe$[9] and is briefly discussed by Bauer in ref. 5 (and references therein). The crossing is similar to what has been hinted to as the effect of a strong magnetic field on the bulk Landau levels of $Bi_2Se_3$ in the appendix of Bernevig and Hughes.[10] This, however, cannot be verified by our measurements.

### 3. Fermi surface geometry in $Pb_{1-x}Sn_xTe$

As shown in Fig. 1(b) and Suppl. Fig. 3, $Pb_{1-x}Sn_xTe$ alloys possess four carrier valleys, one of which – parallel to the [111] direction – is called longitudinal valley. The longitudinal valley is parallel to the growth axis and the direction of the applied magnetic field. The remaining three valleys – referred to as the oblique valleys are tilted by an angle of approximately 70.5° with respect to the [111] direction. When applying a magnetic field in the [111] direction, the Landau level spacing (and the Shubnikov-de-Haas frequency) of each valley is proportional to the area resulting from the cross-section of each respective valley and a (111) plane. For the longitudinal valley this results in a circle of diameter $2a$ - the minor axis of the 3D Fermi ellipsoid. For any tilted valleys, that cross-section is an ellipse of minor axis equal to $2a$ and major axis equal to $2k(\theta)$, where $\theta=70.5°$ is the tilt angle of the oblique valleys when B is parallel to the [111] direction. The area anisotropy factor is $K = \left(\dfrac{b}{a}\right)^2 = 10$ for $Pb_{1-x}Sn_xTe$ when $x=0.40$[11] ($2b$ is the major axis of the 3D ellipsoid). We find for any angle $\theta$:

$$k(\theta) = a\sqrt{\dfrac{K}{(K-1)\cos^2(\theta)+1}} \quad (10)$$

$$k(70.5°) = 2.2a$$

The resulting mass anisotropy between the oblique (m'*) and the longitudinal valleys (m*) is then m'* ≈ 2.2m*. The Fermi velocity anisotropy is then given by $v_f = 1.48 v'_F$.

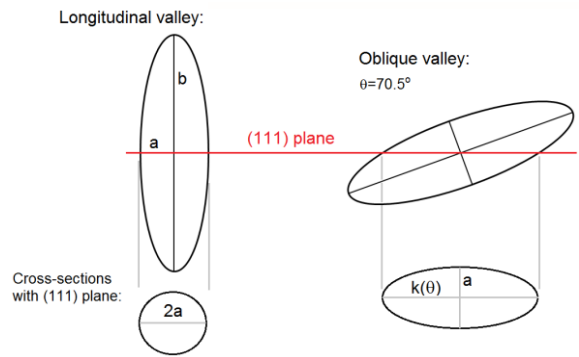

Supplementary Figure 3. Illustration of the Fermi surface geometry. The oblique valley is tilted by 70.5° with respect to the [111] axis and yields an elliptical cross-section with the (111) plane. The figure is not on scale.

References:

1. Dimmock, J. k.p theory for the conduction and valence bands of $Pb_{1-x}Sn_xTe$ and $Pb_{1-x}Sn_xSe$ alloys. in *Proceedings of the International Conference on the Physics of Semimetals and Narrow Gap Semiconductors, 1969, edited by D. L. Carter and R. T. Bate* 319 (Pergamon, New York, 1971).